\documentclass[twocolumn,showpacs,preprintnumbers,prb,epsfig,superscriptaddress]{revtex4}

\usepackage{graphicx}
\usepackage{dcolumn}
\usepackage{bm}

\usepackage{array}  
\usepackage{multirow}  
\usepackage{textcomp} 

\begin{document}

\preprint{K. W. Kim: Sr$_{2}$CuO$_{3}$-II, peak identity}

\title{Optical excitations in Sr$_{2}$CuO$_{3}$}

\author{K. W. Kim}
\email{kyungwan.kim@gmail.com}

\affiliation{Department of Physics, University of Fribourg, Chemin
du Mus$\acute{e}$e 3, CH-1700 Fribourg, Switzerland}

\affiliation{School of Physics and Research Center for Oxide
Electronics, Seoul National University, Seoul 151-747, Korea}


\author{G. D. Gu}
\affiliation{Condensed Matter Physics and Materials Science
Department, Brookhaven National Laboratory, Upton, New York 11973,
USA}


\date{\today}

\begin{abstract}
We investigated excitation spectra of the one-dimensional chain
compound Sr$_{2}$CuO$_{3}$. The small peak at 2.3 eV in the loss
function turned out to correspond to the strong charge transfer
transition at 1.8 eV in conductivity. It has the excitonic
character expected in one dimensional extended Hubbard model of
the transition from the lower Hubbard band to the Zhang-Rice
singlet state. The strongest peak at 2.7 eV in the loss function
is attributed to the continuum excitation of the excitonic charge
transfer transition. The spectral weight sum rule is satisfied
within these transitions.
\end{abstract}


\maketitle

Superconducting cuprates have a CuO$_{2}$ plane in common which is
composed of a two dimensional (2D) corner-sharing network of
CuO$_{4}$ plaquettes. The electronic property of an undoped
CuO$_{2}$ plane should be metallic without strong onsite Coulomb
repulsion at Cu sites, which divides the Cu $d_{x^{2}-y^{2}}$
state into two Hubbard bands of upper Hubbard band (UHB) and lower
Hubbard band (LHB) opening a charge transfer gap between LHB (from
a hole point of a view) and O 2$p$ states. Charge doping changes
the electronic structure of the CuO$_{2}$ plane finally giving
rise to the high temperature superconductivity. It is important to
understand the properties of insulating cuprates to solve the
mystery of the high temperature superconductivity.

Optical spectroscopy is one of fundamental tools to investigate
electronic structure, which can directly measure the charge
transfer gap. By absorbing light, holes in LHB can be excited to O
2$p$ states. When a hole is introduced into the CuO$_{2}$ plane,
the ground state becomes a well known Zhang-Rice singlet (ZRS)
state evenly residing at surrounding O 2$p$ orbitals.\cite{ZR
singlet} Because an optical excitation is a charge conserving
process, to be involved in optical excitations for the ZRS state,
one CuO$_{4}$ plaquette should give a hole to a neighboring
CuO$_{4}$ plaquette, which is possible in corner-sharing CuO$_{4}$
plaquette structures. Therefore, the first optical excitation
crossing the charge transfer gap should be from LHB to the ZRS
state in the CuO$_{2}$ plane as depicted in fig. 1. The second
excitation is expected to be from LHB to nonbonding (NB) O 2$p$
states in the CuO$_{2}$ plane, which is localized within one
plaquette. Energies of these two excitations are essential to
model the electronic structure of the CuO$_{2}$ plane.

However, the electronic structure of insulating cuprates remains
still unclear. Optical spectra of insulating cuprates of 1D and 2D
corner-sharing structure show similar spectral feature of two
peaks around 2 eV like the peaks A and B in fig. 2. It is clear
that the lowest energy peak, which is rather sharp and has a large
spectral weight in common, should be the charge transfer
transition $\alpha$ from LHB to ZRS. But the origin of the other
peak, which comes at about 0.5 eV higher, is still unclear. This
peak is relatively broad and the strength varies depending on
materials. It is rather small and appear as a hump-like shape in
RE$_{2}$CuO$_{4}$ (RE: rare earth ions) and Sr$_{2}$CuO$_{3}$.
\cite{Imada, T-prime 214} Imada \textit{et al}. attributed this
peak to a side band transition with spin excitation via the strong
Kondo coupling between the 2$p$ hole and the 3$d$ spin in the
final state.\cite{Imada} On the other hand this peak has a
comparable or even larger weight than the first peak in
YBa$_{2}$Cu$_{3}$O$_{6}$ and Sr$_{2}$CuO$_{2}$Cl$_{2}$.
\cite{SCOC-Choi, YBCO} H. S. Choi \textit{et al}. attributed this
peak to the $\beta$ transition
from LHB to NB state.\cite{SCOC-Choi} 
Therefore further studies are required to reveal the nature of
those peaks and to understand the electronic structure of
insulating cuprates.

Sr$_{2}$CuO$_{3}$ has a corner-sharing CuO$_{4}$ plaquette 1D
chain structure. Its electronic properties along the chain
direction should have a common character with the CuO$_{2}$ plane,
while only localized phenomena within one plaquette are expected
in the direction perpendicular to the chain. This chain structure
provides a unique chance to investigate the electronic structure
of CuO$_{4}$ plaquette networks telling the difference between
localized and delocalized excitations in a single material. In
addition, the one dimensionality of this material enriches its
physics attracting many researchers with interesting phenomena
such as the spin-charge separation, a large optical nonlinear
effect, and Wannier-like bound excitons.\cite{M. Ono-PRB,
Nonlinearity-Nature, SCO-K. W. Kim, T. Kidd} All these factors
make the excitation spectra of this system of great interest.

\begin{figure}
\includegraphics[width=8cm]{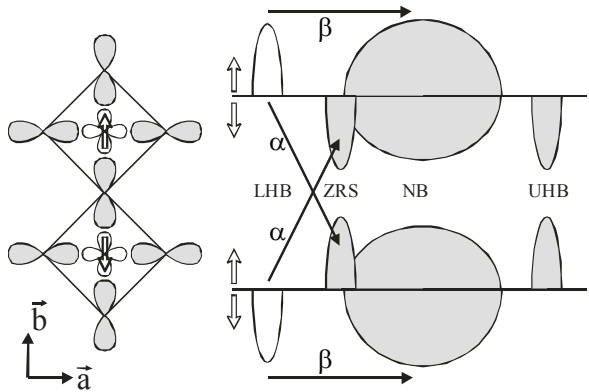}
\caption{Schematic diagram of DOS and optical excitations in a
corner-sharing CuO$_{4}$ plaquette structure with the
antiferromagnetic background from a hole point of view. The greyed
DOS are occupied by electrons. Note that the notations of two
Hubbard bands of LHB and UHB are exchanged in the electron point
of view. The axes of Sr$_{2}$CuO$_{3}$ are given with respect to
the chain.} \label{fig.schematic}
\end{figure}

A. S. Moskvin \textit{et al}. investigated the electronic
structure of Sr$_{2}$CuO$_{3}$ along and perpendicular to the
chain direction by electron energy loss spectroscopy (EELS).
\cite{EELS2} Peaks unique along the chain direction are considered
to be the ZRS state involved transitions, while structures common
in both directions are attributed to transitions localized within
one plaquette. Interestingly, the strong charge transfer
excitation right above the gap was found to be composed of two
peaks. The first small peak above the gap was argued to be a
localized excitation within one plaquette. This observation has
put more importance on the multi-band model accounting details of
orbitals over the simple one band Hubbard model approach.
Unfortunately optical experiment along $a$-axis is lacking. It is
because this material is cleaved along $bc$ plane and its strongly
reactive nature in air makes polishing for a-axis troublesome.
There is only one optical absorption report along $a$-axis
measured on a thin film grown with $ab$ plane utilizing the
anisotropic property of a LaSrAlO$_{4}$
substrate.\cite{film-absorption} Although EELS experiment has
detected an energy loss at 2.0 eV along $a$-axis, no corresponding
absorption feature was observed. This discrepancy demands further
investigations on this system.

In this paper we report optical spectra along the chain direction
of Sr$_{2}$CuO$_{3}$ measured by reflectivity measurement. The
optical conductivity spectra $\sigma(\omega)$ and the loss
function agree with previously reported data. However, we notice
that the observed small peak just above the gap in the loss
function is not a localized excitation within one CuO$_{4}$
plaquette but the ZRS state involved charge transfer excitation
which appears as a strong peak at 1.8 eV in $\sigma(\omega)$. This
peak is an excitonic peak due to the strong inter-site Coulomb
interaction expected in the extended Hubbard model. The strong
peak at 2.7 eV in the loss function is also attributed to the
excitation from LHB to the ZRS state, which is the continuum
excitation of the excitonic peak.

Single crystalline samples were grown using the traveling-solvent
floating zone method. Temperature dependent polarized reflectivity
spectra were carefully measured over a wide energy range. In a low
energy region of 30 to 24000 cm-1 (4 meV to 3 eV) an in situ
evaporation technique was adopted in the overfilling method on
Bruker 66v /S Fourier transform spectrometer. In 4000-50000 cm-1
(0.5 eV to about 6 eV) Cary5 grating spectrometer was used in the
underfilling method. High energy spectra in 6-30 eV were measured
at room temperature utilizing synchrotron radiation from the
normal incidence monochromater beam line at Pohang Light Source
(PLS). All measurements were done on freshly cleaved surfaces.
The complex optical conductivity spectra $\tilde{\sigma}(\omega)$
were obtained from Kramers-Kronig transformation of reflectivity
$\emph{R}(\omega)$.

\begin{figure}
\includegraphics[width=8cm]{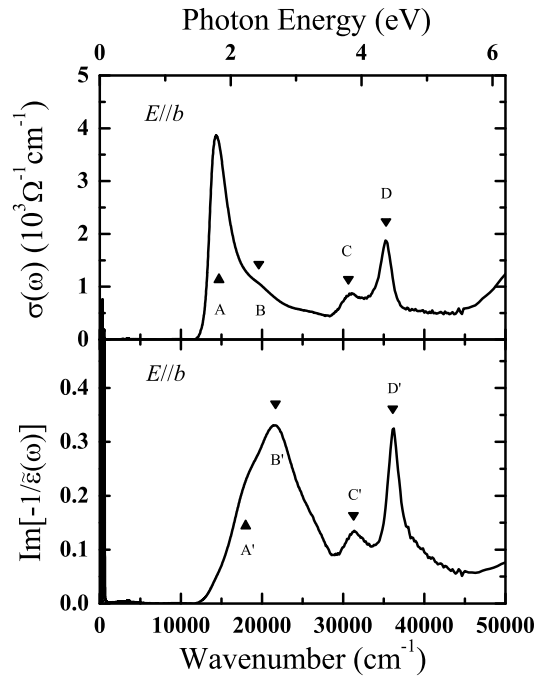}
\caption{Optical conductivity and loss function along the chain
direction of Sr$_{2}$CuO$_{3}$. Four excitations are observed in
both spectra.} \label{fig.cond-loss}
\end{figure}

Figure 2 shows $\sigma(\omega)$ and the loss function along the
chain direction of Sr$_{2}$CuO$_{3}$. Note that the loss function
Im[-1/$\tilde{\epsilon}(\omega)$], which corresponds to EELS
spectrum at zone center, can be obtained also from
$\tilde{\sigma}(\omega) =
\frac{i\omega}{4\pi}(1-\tilde{\epsilon}(\omega))$. There are 4
clear peaks respectively marked as A-D and
A$^{\prime}$-D$^{\prime}$. 
All these peaks were observed in previous studies on
$\sigma(\omega)$ and EELS spectra respectively.\cite{Imada, EELS1,
EELS2} It can be easily noticed that C and D in $\sigma(\omega)$
should correspond to C$^{\prime}$ and D$^{\prime}$ in the loss
function. However coming to A, B and A$^{\prime}$, B$^{\prime}$
their strengths and energies do not allow a simple conclusion.

\begin{figure}
\includegraphics[width=8cm]{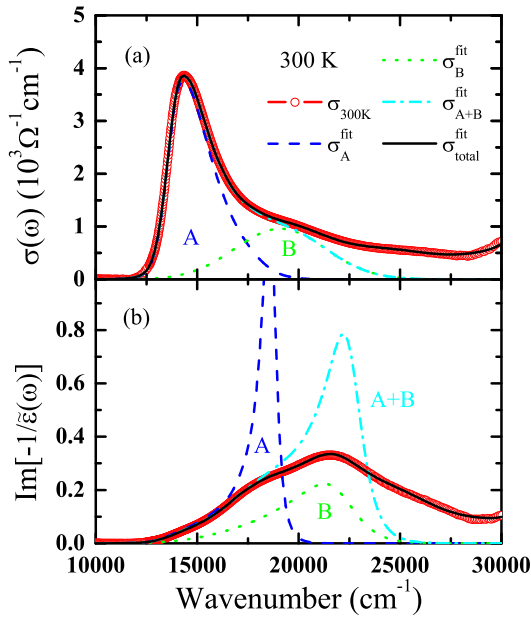}
\caption{(Color online) Gaussian fitting of $\sigma(\omega)$ for
the peak A, B and corresponding loss functions along the chain
direction of Sr$_{2}$CuO$_{3}$.} \label{fig.loss-ftn}
\end{figure}

To find a direct relationship between peaks in $\sigma(\omega)$
and the loss function, $\tilde{\sigma}(\omega)$ was fitted with
Gaussian functions. Note that the strongly asymmetric line shape
of the peak A does not allow Lorentzian with a long tail. Neither
it is possible to fit the peak A with a single Gaussian but at
least 3 Gaussians are required for a decent fitting. Figure 3(a)
shows the fitting result together with an individual peak
corresponding to peaks A and B. The complex dielectric function
$\tilde{\epsilon}(\omega)$ can be decomposed with individual modes
as $\tilde{\epsilon}(\omega) = 1+
\sum_{n}\tilde{\epsilon}_{n}(\omega)$. If all modes are separated
far away from each other then the loss function of each mode can
be obtained simply with its own dielectric function
$\tilde{\epsilon}_{i}(\omega)$ together with
$\epsilon_{\infty}\equiv 1+\sum_{E_{n}>E_{i}}\epsilon_{n}(0)$
which accounts the contribution by higher energy modes. In a real
system, however, some modes are located close to each other such
that the loss functions obtained in this way become slightly
different from peaks observed in the total loss function. That is,
differently from $\sigma(\omega)$ the sum of respective loss
functions obtained in this way does not become the total loss
function. Figure 3(b) shows corresponding loss functions of peaks
A and B obtained as described above. Because two modes A and B are
not completely separated, the obtained loss function for each mode
is different from the respective peaks in the total loss function.
Although their peak heights are much different, however, their
peak energies are well reproduced. Note that the loss function of
the sum of those two modes obtained in the same way already
explains even the peak height quite well for the peak
A$^{\prime}$. This clearly shows that the peaks A$^{\prime}$ and
B$^{\prime}$ correspond to the peaks A and B.

In this corner-sharing chain structure, the first strong peak A
should be the ZRS transition $\alpha$. Note that the strength has
to be considered in $\sigma(\omega)$ not in the loss function.
Then, how can we understand the observed EELS spectra? 
In the reported EELS spectra, there is one more excitation at 5.2
eV which is missing in the loss function shown in fig. 2.
\cite{EELS1, EELS2} Note that the peak C$^{\prime}$ was observed
only with $q\sim 0$. Although the peak at 5.2 eV is missing in our
measurement, the previously reported $\sigma(\omega)$ by Imada
\textit{et al}. also show a small peak there.\cite{Imada} However,
although this peak is common in both spectra along $b$ and $c$
axes, it is much stronger along $c$ axis. This makes it suspicious
that the weak feature along $b$ axis might come from $c$-axis due
to slightly misaligned polarization. Likewise the 2 eV excitation
along $a$ axis in EELS spectra could come from partially mixed $b$
axis.

The low dimensional character of Sr$_{2}$CuO$_{3}$ allows a direct
comparison of the experiment and a theory. The strongly asymmetric
line shape has been understood based on 1D one band extended
Hubbard model (EHM) including on site Coulomb repulsion $U$ and
the nearest neighbor interaction $V$, which has explained many
other properties of 1D charge transfer insulators including
Sr$_{2}$CuO$_{3}$, Ca$_{2}$CuO$_{3}$, and Ni-halogen bridged 1D
materials.\cite{SCO-K. W. Kim, DDMRG-SCO, M. Ono-PRB, EELS1,
film-absorption} However, a close look on $\sigma(\omega)$ of
those 1D charge transfer insulators finds another common feature
which is not pronounced in 1D EHM. That is, like the peak B in
Sr$_{2}$CuO$_{3}$, they have a hump structure in the long tail
above the excitonic peak. Although it does not appear as a
separated peak when $V<2t$, if the excitonic peak becomes a bound
exciton ($V>2t$), the theory expects a separated continuum
excitation above the excitonic peak. The continuum nature of those
peaks was clearly demonstrated in the Ni-Br-Br compound, which
forms a bound exciton at low temperature.\cite{M. Ono-PRL} Note
that the continuum excitation in the Ni-Br-Br compound appears as
a hump even at high temperature, which is clearly manifested in
reflectivity spectra. \cite{M. Ono-PRB} Therefore the peak B could
be attributed to the continuum excitation. That is, if there were
no inter-site Coulomb interaction, the peak B would be the only
transition from LHB to the ZRS state. However, sizable inter-site
Coulomb interaction makes the excitonic peak A take most of the
spectral weight, leaving relatively small weight for the continuum
excitation B.\cite{comment-Maekawa}

This assignment of the peak B contrasts to previous assignments in
insulating cuprates.\cite{Imada, SCOC-Choi} To be assigned as a
spin excitation side band or the $\beta$ transition, the peak B
has to be either weak or strong, which does not hold true among
insulating cuprates. However, if peaks A and B were induced by a
strong excitonic effect, the strength variation could be
understood by different strength of the excitonic effect. Because
the inter-site Coulomb interaction $V$ in 1D EHM lies close to the
critical boundary of $V \sim 2t$ in Sr$_{2}$CuO$_{3}$, a small
lattice and/or chemical environment change may affect the
electronic structure rather strongly.\cite{SCO-K. W. Kim} Even
though the inter-site interaction has been considered as an
important parameter in 1D cases, it has seldom been considered in
2D cuprates for the simplicity in theoretical treatments. 
However, the importance of the excitonic effect in the CuO$_{2}$
plane has been recognized by a simplified calculation on CuO$_{4}$
plaquettes clusters. Hanamura \textit{et al}. elaborated to
explain the discrepancy of the strengths of two peaks A and B
among cuprates by considering the inter-site interaction between
Cu and O. They argued that the A peak is an charge transfer
exciton between Cu $d$ electron and O 2$p$ hole while the B peak
has a continuum character delocalized over a few plaquettes, which
is similar to the assignment in the previous paragraph.
\cite{Hanamura} However, they did not consider the ZRS state but
treated with the O 2$p$ state at each O site for the O 2$p$ hole.
As a result, the excitonic peak A can be obtained even in a single
CuO$_{4}$ plaquette in their calculation, which can not explain
the observed anisotropy between $a$ and $b$ axes of
Sr$_{2}$CuO$_{3}$ and much larger gaps of edge-sharing chain
compounds. \cite{Li-212, comment-Hanamura} Nevertheless their
discussion could be qualitatively valid if a single O 2$p$ state
could stand for the ZRS state at the neighboring plaquette as
Zhang and Ng assumed to explain the dispersion of the excitonic
peak A. \cite{Zhang-Ng} Further theoretical efforts are required
to understand the charge transfer excitations and the excitonic
effect in various insulating cuprates.

Because both peaks A and B have the ZRS transition character, the
next higher energy peaks C and D are expected to be $\beta$
transitions from LHB to NB states. Without spectrum along $a$
axis, it is necessary to rely on general features observed in
insulating cuprates. To investigate the $\beta$ excitation, an
edge sharing chain system such as Li$_{2}$CuO$_{2}$ should be
ideal, because the $\beta$ transition should be the first
pronounced excitation in the edge sharing chain. In
Li$_{2}$CuO$_{2}$ the first strong peak is observed at about 4.2
eV in $\sigma(\omega)$.\cite{Li-212} Recently J. M\'{a}lek
\textit{et al}. did detailed calculation for Li$_{2}$CuO$_{2}$ on
clusters within a 3 band ($pd$) Hubbard model.
\cite{ZRT-contribution} Interestingly, a single plaquette, of
which behavior should correspond to that of the $a$-axis of
Sr$_{2}$CuO$_{3}$, shows two peaks similar to peaks C and D. The
antiferromagnetic background in the corner-sharing chain, which
should suppress the Zhang-Rice triplet (ZRT) transition, could
make these $\beta$ transitions along the chain direction similar
to those of a single plaquette. Note that many 2D insulating
cuprates also have a strong peak in this energy range.
\cite{T-prime 214, Imada, YBCO} Therefore peaks C and D are
attributed to $\beta$ transitions from LHB to NB states.

\begin{figure}
\includegraphics[width=8cm]{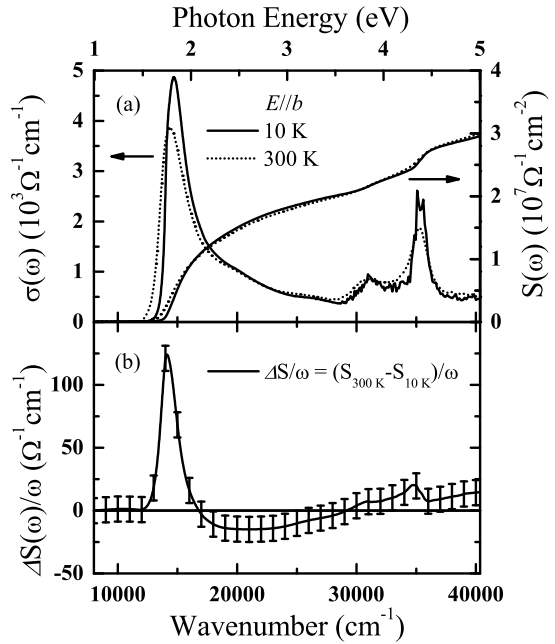}
\caption{(a) Optical conductivity and spectral weight along the
chain direction of Sr$_{2}$CuO$_{3}$ at 300 K and 10 K and (b) its
difference normalized by the energy. The error bar is assumed from
the conductivity value in the gap region. } \label{fig.sw}
\end{figure}

J. M\'{a}lek \textit{et al}. argued also that the spectral weight
of the ZRS excitation at high temperature should decrease due to
the thermally excited ZRT transition, which comes around 4 eV,
even in antiferromagnetic CuO$_{4}$ plaquette systems including
Sr$_{2}$CuO$_{3}$.\cite{ZRT-contribution} Figure 4(a) shows
temperature dependent $\sigma(\omega)$ up to 5 eV and integrated
spectral weight. Because $\sigma(\omega)$ was obtained by
Kramers-Kronig transformation from $\emph{R}(\omega)$, it is
difficult to estimate the error. Only in the gap region, where is
no absorption, the error could be estimated rather easily.
Comparison of the values in the gap region with other reported
data suggests a rather small error in our spectra.\cite{SCO-K. W.
Kim, Imada, M. Ono-PRB} To examine the spectral weight
redistribution, the spectral weight change between 300 K and 10 K
normalized by energy is shown in fig. 4(b). Note that the
normalized spectral weight will have the error similar to that in
$\sigma(\omega)$ because the spectral weight is an integrated
value over energy. There is strong temperature dependence in peak
A, which has been understood by the electron-phonon
coupling.\cite{M. Ono-PRB} The spectral weight sum rule is
satisfied at about 3.5 eV within experimental error. This implies
that the ZRT excitation in this material is absent. Its robust
antiferromagnetic configuration with the largest exchange energy
among cuprates may not allow the ZRT transition which requires
finite ferromagnetic interaction. Because it is difficult to
separate peaks A and B in a consistent way at all temperatures
(note that even the peak A itself requires a few Gaussians for a
decent fitting) it would be meaningless to discuss further details
on the spectral weight. However, it has been noticed that
$\sigma(\omega)$ can be simply scaled by the peak height and width
right above the gap.\cite{SCO-K. W. Kim} Best scaling factors
expect decrease of spectral weight at low temperature. But it
should be noted that the peak A could be composed of a few peaks
due to electron-phonon coupling. \cite{el-ph} In addition, bound
exciton peaks appear at low temperature of which spectral weight
is not taken into account in the scaling argument.\cite{SCO-K. W.
Kim} Therefore, it could be stated that spectral weight of the
bound excitons may come mainly from a lower part of the peak A.
But any further argument is beyond current observations.

In summary, optical conductivity spectra and the loss function of
Sr$_{2}$CuO$_{3}$ were analyzed in detail. Four excitations below
5 eV were attributed to the ZRS state and NB O 2$p$ states
involved excitations respectively. The small peak at 2.3 eV in
EELS spectra, which had been believed to be a localized excitation
within single CuO$_{4}$ plaquette, turned out to correspond to the
strong excitonic charge transfer excitation at 1.8 eV in
$\sigma(\omega)$. Another ZRS state involved transition was
addressed, which appears as the strongest peak at 2.7 eV in the
loss function and corresponds to the continuum excitation in 1D
EHM. The spectral weight sum rule is satisfied within ZRS involved
excitations below 3.5 eV. This finding of the existence of an
excitonic peak and its continuum excitation demands reexamination
of the electronic structure of 2D cuprates.

\begin{acknowledgments}
This work is supported by the Schweizer Nationalfonds (SNF) with
grant 200020- 119784, by the Department of Energy under contract
No. DE-AC02-98CH10886. The experiments at PLS was supported by
MOST and POSCO.

\end{acknowledgments}


\end{document}